\documentstyle[epsf,rotate]{mn}

\newif\ifAMStwofonts
\AMStwofontstrue

\def\hmpc{{\, {\rm h}^{-1}~\rm Mpc}}

\def\eV{{\rm~eV}}

\def\mpc{{\rm~Mpc}}

\def\'{^{\prime}}
\def\avrg#1{{\langle #1 \rangle}}

\def\eps{\varepsilon}

\def\eg{{e.g., }}

\def\etal{{et al. }}

\def\etc{{\frenchspacing etc.}}

\def\half{{\textstyle{1\over2}}}

\def\spose#1{\hbox to 0pt{#1\hss}}
\def\lta{\mathrel{\spose{\lower 3pt\hbox{$\mathchar"218$}}
     \raise 2.0pt\hbox{$\mathchar"13C$}}}
\def\gta{\mathrel{\spose{\lower 3pt\hbox{$\mathchar"218$}}
     \raise 2.0pt\hbox{$\mathchar"13E$}}}
\def\ge{\mathrel{\spose{\lower 3pt\hbox{$-$}}
     \raise 2.0pt\hbox{$\mathchar"13E$}}}
\def\le{\mathrel{\spose{\lower 3pt\hbox{$-$}}
     \raise 2.0pt\hbox{$\mathchar"13C$}}}
\def\simgt{\gta}
\def\simlt{\lta}

\begin{document}

\font\japit = cmti10 at 10truept

\title[Forecasting Cosmic Parameter Errors from the CMB]{Forecasting Cosmic 
Parameter Errors
from Microwave Background Anisotropy Experiments}

\author[Bond, Efstathiou \& Tegmark]{J.R. Bond$^{1}$, G. Efstathiou$^{2}$ and M. Tegmark$^{3,4}$\\
$^{1}$ CIAR Cosmlogy Program, CITA, University of Toronto, Toronto, ON
M5S 3H8, Canada;  bond@cita.utoronto.ca \\
$^{2}$ Department of Physics, University of Oxford, g.efstathiou1@physics.oxford.ac.uk \\
$^{3}$ Institute for Advanced Study, Princeton, NJ 08540, USA; max@ias.edu\\
$^{4}$ Hubble Fellow}

\date{}
\pagerange{}

\voffset -1cm

\label{firstpage}

\maketitle 

\begin{abstract}
Accurate measurements of the cosmic microwave background (CMB)
anisotropies with an angular resolution of a few arcminutes can be
used to determine fundamental cosmological parameters such as 
the densities of baryons, cold and hot dark matter, and certain
combinations of the cosmological
constant and the curvature of the Universe to
percent-level precision. 
Assuming the true theory is a variant of inflationary cold dark matter
cosmologies, we calculate the accuracy with which cosmological
parameters can be determined by the next generation of CMB satellites,
MAP and Planck.  We pay special attention to: (a) the accuracy of the
computed derivatives of the CMB power spectrum ${\rm C}_\ell$; (b) the
number and choices of parameters; (c) the inclusion of prior knowledge
of the values of various parameters.
\end{abstract}

\begin{keywords}
{cosmology: theory --- cosmic background radiation}
\end{keywords}

\section{Introduction} \label{sec:intro}

The detection of primordial anisotropies in the microwave background
radiation by the COBE satellite (Smoot \etal 1992) has had an enormous
impact on cosmology (see White, Scott and Silk 1994 and Bond 1996 for
reviews).  However, the relatively poor angular resolution of COBE/DMR
($\theta_{fwhm} \approx 7^\circ$) limits the amount of information
that can be extracted from the CMB. From the 4 year COBE maps (Bennett
\etal 1996a), the
overall amplitude of the CMB power spectrum for a given spectral shape
has been determined to an accuracy of $7 \%$ and a power law index
characterizing the shape to $\pm 0.24$ (Bond 1996). Constraints
on other parameters such as the spatial curvature and the cosmological
constant $\Lambda$ are weak.

It has long been known (Sunyaev and Zeldovich 1970) that at angular
resolutions smaller than $\sim 1^\circ \Omega_0^{1/2}$ (the angle
subtended by the Hubble radius at the time of recombination) the CMB
power spectrum will depend on \eg the sound speed of the baryon-photon
fluid, and hence on a number of fundamental cosmological parameters,
such as the densities of baryons, cold and hot dark matter, and the
spatial curvature of the Universe. In adiabatic models, the acoustic
motions of the matter radiation fluid lead to a characteristic series
of `Doppler peaks' in the CMB power spectrum which have been
investigated in considerable detail numerically and semi-analytically
(\eg Bond 1996, Hu {\etal} 1997).  Similar behaviour is expected
qualitatively in defect (isocurvature) theories, though the pattern of
Doppler peaks is expected to be less distinct and has not yet been
calculated to high precision (\eg Turok 1996). We therefore restrict
the discussion in this paper to purely adiabatic perturbations obeying
Gaussian statistics, as expected in most inflationary models of the
early universe (\eg Linde 1990). The anisotropies in such models can
be computed to high accuracy which, as we will show in
Section~\ref{sec:param}, is essential for estimating the precision
with which cosmological parameters can be determined from the CMB.

\begin{figure}
\noindent\hskip-0.6cm{
{\vbox{\epsfxsize=10cm
\epsfbox{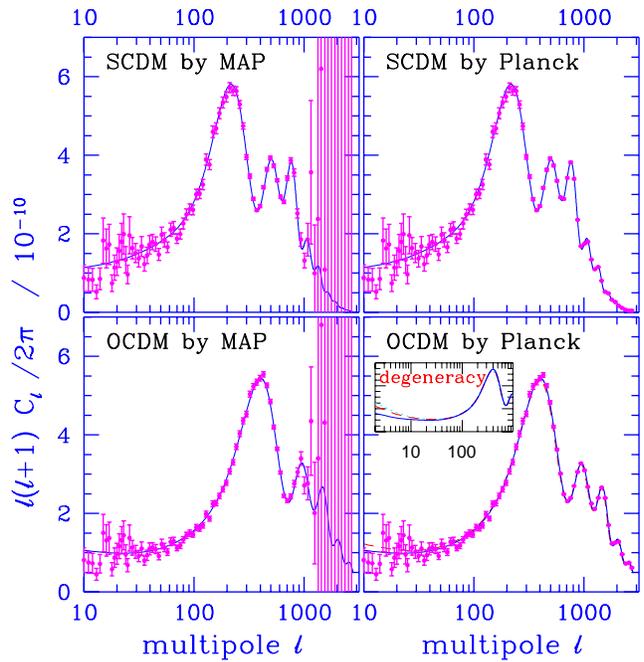}}
}
}
\vskip-0.2cm 
\caption{Temperature power spectra for a standard CDM (SCDM)
model with  ${\rm h}=0.5$, $\Omega_{m}=1$ and
$\Omega_{\Lambda}=0$ and an open CDM (OCDM) model with ${\rm h}=0.6$ ,
$\Omega_{m} = 0.33$ and $\Omega_{\Lambda}=0$. The points show a
quadratic power spectrum estimation of ${\rm C}_\ell$ (in bands of
width 5\% in $\ell$), along with the one sigma error, from simulated
CMB skies observed by the MAP and Planck satellites described in
Section 3 and Table 1. The errors in the left panels use the MAP$^{+}$
parameters for the three highest frequency channels and two years of
observing, while those in the right panels use the four lowest
frequency Planck HFI channels. Planck can follow the theoretical
curves quite precisely far down the damping tail for both models.  In
the bottom right panel, a ${\rm C}_\ell$ curve with $\Omega_m=0.15$,
$\Omega_{\Lambda}=0.44$ and ${\rm h}=0.9$ almost degenerate with the
OCDM model is superposed. The inset, showing these two and also a
$\Omega_m=0.10$, $\Omega_{\Lambda}=0.58$ and ${\rm h}=1.1$ model,
demonstrate that the angle-distance degeneracy relation (see text) used
to define these models is almost exact; these models can be
distinguished only at low multipoles no matter how precise the CMB
experiment.}
\label{fig:CLpower}
\end{figure}

Intermediate angle experiments have detected temperature anisotropies 
which are  consistent with a
primordial origin and in rough agreement with adiabatic theory
predictions. However, the accuracy and robustness of the results does not
yet allow strong conclusions to be drawn, even when experimental
results are combined together ({\eg} Bond 1996, Bond \& Jaffe 1997,
Lineweaver {\etal} 1997, Rocha \& Hancock 1997).  An experiment with
an angular resolution of $\theta_{fwhm}
\sim 5^\prime$ can yield useful information about the CMB
spectrum\footnote{The 
temperature power spectrum is defined as the
expectation value ${\rm C}_\ell = \langle \vert a_{\ell m} \vert^2 \rangle$,
where the coefficients $a_{\ell m}$ are defined by a spherical
harmonic expansion of the temperature anisotropies on the celestial
sphere $\Delta T/ T = \sum_{\ell m} a_{\ell m} Y_{\ell m}( \theta,
\phi)$.} 
${\rm C}_\ell$ up to multipoles beyond a Gaussian filtering scale $\ell_s =
\sqrt{8\ln2}\theta_{fwhm}^{-1} \sim 2000$. If the sky coverage is
complete, each multipole is statistically independent. However, it is
clear from visual inspection ({\it e.g.}
Figure~\ref{fig:CLpower}\footnote{${\rm h}$ is the present value of
the Hubble parameter $H_0$ in units of $100\; {\rm km}\;{\rm s}^{-1}
{\rm Mpc}^{-1}$. The parameters $\Omega_i$ denote the cosmological
densities of various components (defined in Section 2.2) in units of
the critical density.}) that a typical inflationary ${\rm C}_\ell$
curve is smooth and can be specified accurately by many fewer than
$2000$ parameters.  It is therefore not obvious {\it a priori} to what
extent the typically 10-15 parameters specifying an inflationary model
can be disentangled by a particular set of measurements. Evidently, a
detailed calculation is necessary as has been performed in an
important paper by Jungman \etal (1996). However, as described in the
next section, an accurate assessment of the degeneracies between
cosmological parameters imposed by high resolution CMB experiments
requires a precise numerical calculation, rather than the
semi-analytic approach adopted by Jungman \etal

\section{Parameter estimation with priors}  \label{sec:param}

\subsection{The covariance matrix}\label{sec:covmat}

Errors on a set of cosmological parameters ${\bf s}=\{ s_i
\},i=1,\ldots ,n $ are estimated using Bayes theorem, which updates
the prior probability $P({\bf s}\vert {\rm prior})$ for the parameters
with a likelihood function ${\cal L}({\bf s})$: $P({\bf s}) \propto
{\cal L}({\bf s}) P({\bf s}\vert {\rm prior})$. (Although a uniform
prior seems to be the least prejudicial, there are certain
non-debatable restrictions and plausible constraints which are
reasonable to include as prior information, as discussed in
Section~\ref{sec:prior}.) If the errors $\delta {\bf s}$ $ \equiv {\bf
s} - {\bf s_\circ} $ about the mean ${\bf s_\circ} = \langle {\bf s} \rangle$
are small, then an expansion of $\ln{\cal L}$
to quadratic order
about the maximum gives:\footnote{The likelihood also has terms which
depend upon the deviations of the specific realization of the theory
represented by the observations from the ensemble-averaged values used
here (Bond 1996). We have tested these effects in detail on parameter
estimation and find them to be small and fully consistent with the
error estimates we quote in the Tables.}
\begin{eqnarray}
{\cal L} &\approx& {\cal L}_m \exp\left[-\half \sum_{ij}{F}_{ij}
\delta s_i \delta s_j\right]\, , \nonumber \\
{F}_{ij} 
&=& \sum_\ell (\Delta {\rm C}_{\ell})^{-2} 
{\partial {\rm C}_{\ell} \over \partial s_i} 
{\partial {\rm C}_{\ell} \over \partial s_j} \, , \label{eq:fishmat} \\
(\Delta {\rm C}_{\ell})^2 &\approx& 
{2\over (2\ell +1)f_{sky} }\, 
\left( {\rm C}_{\ell} 
+ \overline{w}^{-1} \overline{{\cal B}}_\ell^{-2}\right)^2 \, ,
\label{eq:Dcell} \\
\overline{w} &\equiv& \sum_c w_c, \quad \overline{{\cal B}}_\ell^2 
\equiv 
 \sum_{c} {\cal B}^2_{c\ell}w_c/\overline{w} , 
\label{eq:chDcell} \\
w_c &\equiv& (\sigma_{c,pix}\theta_{c,pix})^{-2}, \quad {\cal B}^2_{c\ell}
\approx e^{-\ell(\ell+1) /\ell_s^2},   \nonumber 
\end{eqnarray} 
adopting a Gaussian approximation for the beam function ${\cal
B}_{c\ell}$.  In this approximation, with a uniform prior the
covariance matrix ${\bf M} \equiv
\langle \delta{\bf s} \delta {\bf s}^\dagger \rangle$ 
is the inverse of the Fisher information matrix ${\bf F}$ (\eg Tegmark
{\etal} 1997), the 1-sigma error on $s_i$ is $\sigma_i =
M_{ii}^{1/2}$, and the
correlation coefficient between parameters $i$ and $j$ is $r_{ij} =
M_{ij}/(\sigma_i \sigma_j)$.  Equation~(\ref{eq:Dcell}) gives the
standard error $\Delta {\rm C}_\ell$ on the estimate of ${\rm C}_\ell$ for an
experiment with $N$ frequency channels $c$, 
angular resolution $\theta_{c,fwhm}$, and sensitivity $\sigma_{c,pix}$
per resolution element ($\theta_{c,fwhm} \times \theta_{c,fwhm}$
pixel), which samples a fraction $f_{\rm sky}$ of the sky. 

A key simplifying assumption in deriving equation~(\ref{eq:Dcell}) is
that the noise is homogeneous and isotropic. Provided that the
sampling varies slowly, accurate error estimates can be obtained using
the average weight for the experiment.  Equation~(\ref{eq:Dcell}) is
exact if $f_{sky} = 1$ (\eg Knox 1995) and is approximately correct at
multipoles $\ell \gta \ell_{cut}$ corresponding to angular scales
small compared to the dimensions $2\pi /\ell_{cut}$ of an incomplete
sky map.

The accuracy of cosmological parameter estimation via the covariance
matrix approach depends on: (1) the validity of the Gaussian
approximation to the likelihood function; (2) the number and choice of
the parameters ${\bf s}$ defining the theoretical model; (3) the
parameters ${\bf s_\circ}$ of the target model; (4) the numerical accuracy
of the derivatives of ${\rm C}_\ell$; (5) the inclusion of prior
constraints on the parameters ${\bf s}$; (6) systematic errors in
estimates of ${\rm C}_\ell$ caused by Galactic and extragalactic
foregrounds.  For high resolution experiments which tightly constrain
many of the cosmological parameters, a Gaussian approximation about
the maximum likelihood should be quite good (Knox 1995, Spergel
private communication), although positivity and other constraints can
truncate general excursions in the likelihood space. We ignore
systematic errors caused by foreground subtraction, since over much of
the sky these are very likely to be much smaller than the variance of
equation~(\ref{eq:Dcell}) (see {\it e.g.} Tegmark and Efstathiou 1996
for a discussion of foreground removal from CMB maps). Here we
consider the remaining four points.

\subsection{Choice of variables} \label{sec:varchoice}

Parameters describing the theoretical angular power spectra include
those for initial conditions and those characterizing the transport of
radiation through photon decoupling to the present. If we were to
allow all possible variations, the count of parameters could easily
exceed $20$; in our analysis, we use $\le 11$ variables.  We
characterize the initial fluctuation spectra by an amplitude and a
spectral index (tilt) for the scalar and tensor components, ${\cal
P}_{\Phi}^{1/2}(k_n) $ and $n_s$, ${\cal P}_{GW}^{1/2}(k_n) $ and
$n_t$. The primordial amplitude parameters for the gravitational
potential fluctuations, ${\cal P}_{\Phi}^{1/2}(k_n) $, and the gravity
wave fluctuations, ${\cal P}_{GW}^{1/2}(k_n) $, are chosen here to be
normalized at a wavenumber corresponding to the horizon scale. The
fluctuations arising from inflation could be much more complicated,
requiring, for example, the parameterization of variations of the
spectral indices with wavenumber $k$, inclusion of isocurvature as
well as adiabatic components in the scalar perturbations, and possibly
of non-Gaussian features. 

At the time of decoupling, the key parameters determining the
temperature power spectrum are the densities of various types of
matter, the expansion rate, the sound speed, and the damping rate; all
of these depend only on the density parameters $\omega_j \equiv
\Omega_j {\rm h}^2$, where $j=b, cdm, hdm, \gamma , er\nu$ refers to
baryons, cold dark matter, hot dark matter, and the various
relativistic particles present then, such as photons and relativistic
neutrinos. The Hubble parameter 
at that time only depends upon
$\omega_{m}=\omega_b+\omega_{cdm}+\omega_{hdm}$ (if the massive
neutrinos were nonrelativistic then\footnote{Massive neutrinos become
nonrelativistic below a redshift $\sim 1700 (m_\nu /{\rm ev})$,
where $\omega_{hdm} \approx 0.01 (m_\nu /{\rm ev}) N_{m\nu}$,
$N_{m\nu}$ is the number of neutrino species of mass $m_\nu$,; for
small ${m_\nu}$, the neutrinos may be relativistic or semirelativistic
at decoupling.}) and $\omega_{er}=
\omega_{\gamma}+\omega_{er\nu}$. 

The transport to an angular structure of scale $\ell^{-1}$ now from
the post-decoupling spatial pattern of temperature fluctuations of
comoving scale $k^{-1}$ depends on 
the cosmological angle-distance relation, $\ell \sim k {\cal R}$,
where
\begin{eqnarray}
{\omega_{m}^{1/2}{\cal R}\over 3000 \mpc}  &=& 
{\omega_{m}^{1/2} \over \omega_k^{1/2}} {\rm sinh}\left[\int
{\omega_k^{1/2} da \over
(\omega_ka^{2}+\omega_{\Lambda}a^{4}+\omega_{m}a)^{1/2}}
\right] \nonumber 
\end{eqnarray}
for an open universe (\eg Bond \& Efstathiou 1984). Here  $\omega_{\Lambda} \equiv
\Omega_{\Lambda} {\rm h}^2$ parameterizes the energy density associated with 
a cosmological constant $\Lambda$ ($\Omega_{\Lambda}=\Lambda
/(3H_0^2)$) and $\omega_k \equiv (1-\Omega_0){\rm h}^2$ =
$(1-\Omega_m-\Omega_\Lambda){\rm h}^2$ parameterizes the energy
associated with the mean curvature of the universe. This results in a
degeneracy along ${\delta (\omega_{m}^{1/2} {\cal R})} = 0$ lines,
which leads to a linear relation between $\delta
\omega_k$ and $\delta \omega_{\Lambda}$ for fixed $\omega_{m}$, with coefficients
that depend upon the explicit target model.  The angular
pattern we observe also depends upon the change of the gravitational
metric in time between post-decoupling and the present, which breaks
this degeneracy.  However, this late-time integrated Sachs-Wolfe
effect influences only low multipoles which have a large cosmic
variance. Thus, there exists one combination of variables which cannot
be determined accurately from CMB observations alone, even with a high
precision experiment such as the Planck Surveyor, as the lower right
panel of Fig.~\ref{fig:CLpower} illustrates. 


Some parameters are tightly constrained by measurements other than CMB
anisotropies. For example, $\omega_\gamma$ depends on the temperature
$T_0$ of the CMB, $\omega_{er\nu}$ depends as well on the number of
massless neutrino types; the ${\rm C}_\ell$'s also depend on the
helium abundance, parameterized by $Y_{He}$.  Rather than allow such
parameters complete freedom, we use the prior probabilities to restrict
their allowed variations. (Since the experimental errors on $Y_{He}$,
$T_0$ and $N_\nu$ are small
\footnote{ $Y_{He} = 0.23 \pm 0.01$ (Pagel \etal 1992), $T_0 = 2.728
\pm 0.004$ (Fixsen \etal 1997), $N_\nu = 2.991 \pm 0.016$ (LEP Electroweak
Working Group 1995).},
they have a weak effect on
other cosmological parameters and hence we include only $Y_{He}$ in 
our analysis to illustrate the methodology.)

There also could be many parameters needed to characterize the
ionization history of the Universe; here we use the Compton optical
depth $\tau_C$ from the present to the redshift of reheating, assuming
full ionization. We therefore analyse a maximum of 11 parameters in
this paper: $Y_{He}$, $\tau_C$, 4 initial condition parameters and 5
density parameters $\omega_j$.

For a given model, the amplitudes of the scalar and tensor power
spectra are uniquely related to the observed amplitude of the CMB
power spectrum and that of the present day mass fluctuations
(characterised, for example, by the {\it rms} density fluctuation in
spheres of radius $8\hmpc$, $\sigma_8$).  For example, Jungman \etal
used the quadrupole ${\rm C}_2^{1/2}$ to fix the amplitude of the
fluctuation spectrum.  We use $\avrg{\ell (\ell +1){\rm C}_\ell
/(2\pi)}_B^{1/2}$, an average over the total band $B$ of multipoles
that is accessible to the experiment, since this is most accurately
determined. However the normalization parameters $\sigma_8$ and ${\cal
P}_\Phi^{1/2}(k_n)$ are of sufficient interest that we also show the
accuracies with which these can be determined. To characterize the
tensor amplitude, we use $r_{ts}\equiv {\rm C}^{(T)}_2/{\rm
C}^{(S)}_2$ instead of ${\cal P}_{GW}^{1/2}(k_n)$.  In inflation
models, ${\cal P}_{GW}^{1/2}(k_n) /{\cal P}_{\Phi}^{1/2}(k_n) $ is
simply related to the tensor tilt, with small corrections dependent
upon the scalar and tensor tilts, so one of the four initial
fluctuation parameters is a function of the other three, here chosen
to be $n_t$. $r_{ts}$ also depends upon other cosmological parameters
as well as the tilts (\eg Bond 1996, equation (6.38)).

Any parameter set which defines a coordinate system on the likelihood
surface is a viable set. However, parameters for which the Fisher
matrix analysis is particularly well suited are those for which the
first order expansion ${\rm C}_\ell = {\rm C}_\ell ({\bf s_\circ}) +
(\partial {\rm C}_\ell({\bf s_\circ})/\partial {\bf s})\cdot ({\bf s}
- {\bf s_\circ})$ is more accurate than the sampling variance $\Delta
{\rm C}_\ell$ for parameters ${\bf s}$ that lie within a  few standard
deviations from the target set ${\bf s_\circ}$.  The set
of variables that we have adopted gives acceptably high accuracy for
the CMB experiments described in Section~\ref{sec:satparams}.

\subsection{Choice of target models}\label{sec:targetmod}

We analyze three spatially flat ($\Omega_k=0$) target models and one
with negative curvature. For the canonical flat universe we use a
standard CDM model (SCDM) with the following parameters: $n_s = 1$,
$n_t = 0$, $\Omega_{m} =1.0$, $\Omega_b = 0.05$, $r_{ts}=0$,
$\Omega_{cdm} = 0.95$, $\Omega_{hdm} = 0$, ${\rm h}=0.5$, $\tau_C =
0$, $Y_{He} = 0.23$; it has $\sigma_8=1.2$, normalized to match COBE
DMR, but with too many clusters to match the observations. The open
model has ${\rm h}=0.6$, $\Omega_m =0.33$, $\Omega_b=0.035$,
$\Omega_{cdm} = 0.30$ and $\sigma_8=0.44$, the COBE-normalization,
which has too few clusters. We also discuss results for two other
DMR-normalized spatially flat models that more closely match
observations: an HCDM model with 2 species of massive neutrinos,
$\Omega_{hdm}=0.2, m_\nu =2.4 \, \eV$, ${\rm h}=0.5$ and
$\sigma_8=0.77$; a $\Lambda$CDM model with $\Omega_\Lambda =0.67$,
${\rm h}=0.7$ and $\sigma_8=1.1$. (All models have a 13 Gyr
cosmological age.)

\subsection{Accuracy of the power spectrum derivatives}\label{sec:accuracy}

Computational errors in the derivatives of ${\rm C}_\ell$ can lead to
large errors in the covariance matrix. We distinguish between two
classes of error, one caused by inadequate semi-analytic
approximations to the ${\rm C}_\ell$ and the second caused by
numerical errors in ${\rm C}_\ell$ and its derivatives computed from
linear Boltzmann transport codes.  The ${\rm C}_\ell$ accuracy must be
$1\%$ or better, especially for high resolution experiments probing
multipoles $\ell \simgt 1000$ where the expected random errors on each
individual multipole become $\simlt 3 \%$. Errors which are weakly
correlated with physical parameters are particularly serious since
these can artificially break real near-degeneracies between
cosmological parameters and lead to overly optimistic error
estimates, sometimes by an order of magnitude or more. Extreme care is
therefore required in computing the ${\rm C}_\ell$ derivatives.
For this work we have used derivatives calculated with two Boltzmann
transport codes, an updated version of the multipole code described by
Bond and Efstathiou (1987) generalized to low density universes and
including tensor components (Bond 1996 and references therein), and
the fast path-history code developed recently by Seljak and
Zaldarriaga (1996). Generally the ${\rm C}_\ell$'s from these codes
agree to better than $1 \%$.  We use intervals of typically $1 - 5 \%$
in the parameters ${\bf s}$ in computing numerical derivatives of
${\rm C}_\ell$, {\it i.e.} small enough that the derivatives are
insensitive to the size of the interval, but large enough that they
are unaffected by numerical errors in the ${\rm C}_\ell$ coefficients.
The primary limitation on the error estimates for Table
\ref{tab:satparams1} should be  the ${\rm C}_\ell$ 
linearization assumption made in deriving equation~\ref{eq:fishmat},
although we believe that better than percent level accuracy in ${\rm
C}_\ell$ is needed to achieve high precision in the nearly degenerate
directions of parameter space.  The differences between our error
estimates and analogous results of Jungman \etal, which are large for
some parameters, are caused primarily by their use of semi-analytic
approximations to calculate ${\rm C}_\ell$ and its
derivatives. Zaldarriaga \etal 1997 have undertaken a similar analysis
and come to similar conclusions as those presented here. They also
showed that polarization information can improve the accuracy of some
variables, \eg $\tau_C$, if foregrounds are ignored. Little is known
about how the polarization of foregrounds will compromise the
relatively weak polarization signal of primary anisotropies,
especially at low $\ell$ where much of the improvement comes from. See
also related work on forecasting errors by Mageuijo and Hobson (1997).

\subsection{Inclusion of prior information on parameters}\label{sec:prior}

We have mentioned that a non-uniform $P({\bf s}\vert {\rm prior})$ is
particularly useful for parameters such as $Y_{He}$ and $T_0$, where
other experiments restrict their values to much higher accuracy than
can be achieved from CMB experiments alone.  For other parameters,
{\it e.g.} $\omega_b$ and $\omega_{cdm}$, it may be that the
distribution derived from a CMB experiment is much narrower than any
reasonable prior distribution, in which case we gain little by
including prior information.  There are also intermediate cases where
prior information can help break degeneracies between parameters
estimated from the CMB alone.  We approximate the prior distribution
of parameter values by a Gaussian distribution with covariance matrix
${\bf T}$, so the covariance matrix of parameter values including
prior information is given by
\begin{eqnarray}
&& {\bf M}  = ({\bf F} + {\bf T}^{-1})^{-1}  \  {\rm if} \ P({\bf s} \vert {\rm
prior} ) \propto e^{-{\half \delta {\bf s}^\dagger {\bf T}^{-1} \delta {\bf
s}}} \label{eq:MFT}
\end{eqnarray}
({\it e.g.} Knox 1995), where ${\bf F}$ is the Fisher matrix
(\ref{eq:fishmat}).  If we are interested in the error bars on $s_i$
irrespective of the values of the other variables, we would
marginalize over these, with error $\sigma_i = M_{ii}^{1/2}$ for the
Gaussian case. For most
of the entries in Table~\ref{tab:satparams1} 
we use no prior at all (`no$P$'), except for $Y_{He}$ where
indicated. When priors are used, we adopted a diagonal covariance
matrix $T_{ij}$ with 
the following values for $\sqrt{T_{ii}}$: 0.3
on the normalization $\delta \avrg{{\cal C}_\ell}_B^{1/2}/
\avrg{{\cal C}_\ell}_B^{1/2}$, 0.5 on $n_s$, 2 on $r_{ts}$, 0.075 on
$\omega_b$, 1 on $\omega_{m}$, 1 on $\omega_{\Lambda}$, 1 on
$\omega_{k}$, 0.5 on $\omega_{hdm}$, and 1 on $\tau_C$.  Some
variables are restricted for physical reasons to lie within a certain
range \eg $\tau_C$ and $r_{ts}$ must be positive. Such constraints can
be incorporated into the prior, but at the expense of more complicated
expressions after marginalization over these constrained variables. In
some cases, imposing physical restrictions can lead to a factor of two
or more improvement in the accuracy of the parameter estimates.

Generally the errors in the parameters will be correlated through
nondiagonal components of $({\bf F}+{\bf T}^{-1})^{-1}$. 
Linear combinations of the parameters which are
uncorrelated can be found by diagonalizing  $({\bf F}+{\bf T}^{-1})$.
When the eigenvalues of $({\bf F}+{\bf T}^{-1})$ are rank ordered,
from highest to lowest, the variable combinations corresponding to
high values will be very accurately determined, while those for the
lowest may be very poorly determined, representing the most degenerate
directions in parameter space. In Tables 2 and 3 below we list the
number of parameter combinations that are determined within a $\pm
0.01$ and $\pm 0.1$ accuracy.

\section{Cosmological Parameter Errors from MAP and Planck} 
\label{sec:satparams}

\subsection{The CMB power spectrum estimated from 
MAP and Planck} \label{sec:pspec}

In this section we apply the above machinery to determine the accuracy
of cosmological parameter estimation from two satellite experiments:
the MAP satellite selected by NASA (Bennett \etal 1996b) and the Planck
Surveyor Mission (formerly named COBRAS/SAMBA) selected by ESA
(Bersanelli \etal 1996). These satellites offer examples of the best
that is likely to be achieved in the next decade.  Ground
based and balloon borne experiments will certainly continue to provide
improved constraints on cosmological parameters over this timescale,
and so we also analyze a sample long duration balloon experiment (LDB).

The specifications adopted for MAP and Planck are listed in Table 1
and have been computed from the information provided on the respective
WWW pages for the two missions. Although indicative of the expected
performance of each satellite at the time of writing, these are likely
to evolve.  Of the 5 HEMT channels for MAP, we assume that the 3
highest frequency channels, at 40, 60 and 90 GHz, will be dominated by
the primary cosmological signal.
We also present the gains that result from a 25\% improvement in
angular resolution at all frequencies and 2 years of observing time
(we denote these specifications by MAP$^{+}$). Such an improvement is
now expected for MAP (Page, private communication).  Planck will have
two detector arrays, a Low Frequency Instrument (LFI) using HEMTs and
a High Frequency Instrument (HFI) using bolometers.  The current
design of the HFI incorporates an additional channel at 100~GHz in
addition to channels at 150, 217 and 353~GHz; we have adopted
parameters as listed in Table 2 for these four channels.  We also
present results for the 3 highest resolution channels in the current
design of the Planck LFI which has an expected performance that is
significantly improved over those given by Bersanelli \etal (1996).

For each multipole $\ell$, the computational procedure automatically
rotates the channels into a linear combination optimal for the CMB. In
practice, a more sophisticated treatment would be required in practice
to remove Galactic and extragalactic foregrounds. It is beyond the
scope of this letter to assess the systematic errors in parameter
estimates arising from inaccurate foreground subtraction. We therefore
simply assume that Galactic foregrounds are negligible over a fraction
of the sky $f_{sky}=0.65$, similar to the `clean' sky area adopted in
most analyses of the COBE power spectrum.

Figure~\ref{fig:CLpower} shows examples of ${\rm C}_\ell$ estimates
from one realization of the SCDM and OCDM target models. In this
figure, the estimated power spectra have been averaged over $5\%$ wide
bands in $\ell$. At the resolution of MAP there is very little useful
information beyond $\ell \sim 800$ (the third acoustic peak for the
spatially flat models), whereas Planck samples the power-spectrum at
close to the theoretical variance limit to multipoles $\ell \sim
2500$.  The consequences of these differences for CMB parameter
estimation are described in the next section.

\subsection{Accuracy of cosmological parameters} \label{sec:numbers}

Results of the analysis for the sample LDB experiment and for the MAP
and Planck satellites are given in Table~\ref{tab:satparams1}. 
For the LDB example we adopt specifications for TopHat (Meyer \etal
1997), which would cover 4.3\% of the sky with an error of 18$\mu$K
per $20^\prime$ pixel (which includes an allowance for the extra error
incurred in removing foregrounds).  We assume 65\% of this area will
be usable. Because the sky coverage is so limited, COBE's DMR is added
to improve the baseline in $\ell$ covered and thus the accuracy of
parameter estimation. (When we apply our analysis to DMR alone, using
the average noise in the 53+90+31 GHz map,
$\overline{\sigma}_{pix}\approx 30\mu K$ per 5.2$^\circ$ pixel, we
predict the bandpower would be determined to 0.07 and $n_s$ to 0.20
for SCDM, 0.07 and 0.28 for OCDM, if only these two parameters are
used, in agreement with what is actually found (Section 1); with all 9
parameters, the errors grow, but the first and second eigenparameter
combinations have error estimates of 0.09 and 0.20 for SCDM.) When
treating two experiments at very different scales, such as DMR and the
long duration balloon experiment example shown, the log likelihoods
(and Fisher matrices) just add.  Instead of TopHat, we could have
chosen any of the other bolometer-based LDBs, such as Boomerang (Lange
\etal 1997) or MAXIMA (Richards \etal 1997) or even HEMT-based
LDBs, such as BEAST (Lubin \etal 1997) and derived similar error forecasts.

  For most parameters the inclusion of prior constraints on their
variation have no effect, particularly for a high precision experiment
like Planck. Even for MAP the inclusion of priors has little impact,
except for variables such as $Y_{He}$ which are poorly constrained
from the experiment alone. If the helium abundance is allowed to float
freely, it has a substantial effect on the other parameters; however,
limiting its value to be $0.23 \pm 0.02$ results in little impact on
the other numbers. 
For the LDB+DMR case, the errors are susbstantially larger with no
controlling priors.

 As expected, the parameters have correlations among themselves that
range from weak to very strong in all models, and can differ from
experiment to experiment as well as model to model. The power
amplitude $\avrg{{\cal C}_\ell}_B^{1/2}$ and $\tau_C$ have a
correlation coefficient about 90\% for SCDM for Planck and MAP. The
most highly correlated are $\omega_{k}$ and $\omega_{\Lambda}$, as
expected from the angle-distance near-degeneracy. In the Tables, the
$\Omega_{\Lambda}{\rm h}^2$ numbers are determined with $\Omega_k{\rm
h}^2$ fixed, and the $\Omega_k{\rm h}^2$ numbers are determined with
$\Omega_{\Lambda}{\rm h}^2$ fixed; the other parameters are relatively
insensitive to fixing either, or neither. Thus, although our estimates
of errors after marginalization are gratifyingly small for many
parameters, especially for the specifications of Planck, they are
large in other cases ({\it e.g.}  $\delta \Omega_{\Lambda}{\rm
h}^2$). Error estimates in square brackets are those obtained when the
most correlated component for that variable is constrained to be the
target value. A more natural way to deal with strong correlations
between variables is to perform a principle component analysis in
parameter space, rank-ordering linear combinations of parameters, as
described in Section~\ref{sec:prior}: some linear combinations are
determined exquisitely well and some are less well determined because
of near-degeneracies as is illustrated in the Tables. The Tables also
show values obtained in round brackets when positivity constraints on
parameters such as $\tau_C$ are used.

The $\delta {\rm h}/ {\rm h}$ shown are determined from ${\rm h}^2 =
\sum_j \omega_j$, hence it is a derived rather than fundamental quantity.
However, ${\rm h}$ errors depend upon what is kept fixed and what is
varied. Thus we can use ${\rm h}$ to replace one of $\omega_{m}$,
$\omega_{\Lambda}$, $\omega_k$, with the other two to be marginalized. In
that case, the error estimate would be $\delta {\rm h}/{\rm h} = 0.5
\delta \omega_j /{\rm h}^2$.

We find that the estimated errors on parameters are sensitive to their
input target-model values. Table~\ref{tab:satparams2} shows results
for two other $\Omega_k=0$ models, a $\Lambda$CDM model and an HCDM
model. This illustrates the sensitivity of parameter error estimation
to relatively modest changes in the target ${\rm C}_\ell$. In
interpreting these tables it is also important to take into account
the restrictions that we have imposed on the models. The OCDM model
error estimates are derived assuming there is no tensor
component. Including it has little effect on the results: even the
most correlated, the amplitude, $\tau_C$ and $n_s$, are only slightly
affected. The angle-distance scaling ensures that the tensor power
spectrum does not fall off until higher $\ell$ than in the
$\Omega_k=0$ cases, and this leads to a substantial improvement in
$\delta r_{ts}$.

We have also found that the errors on $ n_t$ and $ r_{ts}$ are
extremely dependent on the input $r_{ts}$ if they are allowed to vary
independently (Knox and Turner 1994, Knox 1995, Efstathiou 1997).
However, ${\rm h}$ and the various matter densities, $\omega_{cdm}$
\etc, are insensitive to the tensor spectrum for reasonable values of
$r_{ts} \simlt 2$.  In open universes, features in the power-spectrum
shift to larger multipoles according to the angle-distance relation,
roughly as $C(\ell)
\rightarrow C(\ell/\Omega_0^{1/2})$; thus, for low $\Omega_0$, 
high resolution is required to determine parameters which affect the
Doppler peak structure ({\it e.g.} ${\rm h}$ and the various
$\omega$).  The relative accuracies of the parameters are less
sensitive to variations of $\Omega_b$ and ${\rm h}$.

\section{Conclusions} \label{sec:conclusions}

In summary, we have described how to compute the errors in the
estimation of cosmological parameters from measurements of the
CMB power spectrum at a number of frequencies with different
angular resolutions and sensitivities. We have also shown how
prior information on the values of parameters can be incorporated
into the analysis and described some of the pitfalls of this type
of analysis that can arise if inaccurate derivatives of the ${\rm C}_\ell$'s
are used and if poor parameter choices are adopted.

We have applied our machinery to the MAP and Planck satellites and
find that these missions are capable of determining fundamental
cosmological parameters to an accuracy that far exceeds that from 
conventional astronomical techniques. In particular, Planck is
capable of determining the Hubble constant and the baryon density
parameter $\Omega_b$ to a precision of a few percent or better for
each of the target models listed in Tables~\ref{tab:satparams1} and
\ref{tab:satparams2}. However, some parameter combinations are poorly
determined by CMB observations alone as described in
Section~\ref{sec:targetmod} and Section~\ref{sec:satparams}. 
Nevertheless, despite this caveat, it is evident from this work that 
accurate CMB observations
have the potential to revolutionize our knowledge of the key cosmological
parameters describing  our Universe.

\smallskip
We would like to thank Lloyd Knox for useful discussions.  JRB was
supported by the Canadian Institute for Advanced Research and
NSERC. GPE acknowledges the award of a PPARC Senior Research
Fellowship.  MT was supported by NASA through a Hubble Fellowship,
{\#}HF-01084.01-96A, awarded by the Space Telescope Science Institute,
which is operated by AURA, Inc. under NASA contract NAS5-26555.

\def\prd{{Phys.~Rev.~D}}
\def\prl{{Phys.~Rev.~Lett.}}
\def\apj{{Ap.~J.}}
\def\apjl{{Ap.~J.~Lett.}}
\def\apjsuppl{{Ap.~J.~Supp.}}
\def\mnras{{M.N.R.A.S.}}

\begin{table*}
\begin{minipage}{120mm}
\caption{Experimental parameters adopted for this study. $\nu_{ch}$ denotes the central
frequency of the channel, $\theta_{fwhm}$ the resolution,
$\sigma_{pix}$ the pixel
sensitivity (in $\Delta T/T$) per $\theta_{fwhm}^2 $ resolution
element, $\ell_s$ is the Gaussian beam filter scale, and $w^{-1}$ is
the noise power for each channel.}
\label{tab:exptparam}
\begin{tabular}{|l|llllll|}
\hline
\multicolumn{7}{c} \hfill MAP (first 3 used) \hfill\\
\hline
$\nu_{ch}$(GHz) & 90 & 60 & 40 & (30) & (22) & \\
$\theta_{fwhm}$ & $18^\prime$ & $23^\prime$ & $32^\prime$ & ($39^\prime$)& ($54^\prime$)& \\
$\sigma_{pix}/10^{-6}$ & 13 & 9.9 & 7.3 & (6) & (4) &  \\
$w^{-1}/10^{-15}$ & $4.5$  & $4.5$ & $4.5$  &  & & \\
$\ell_s$ & 465 & 345 & 255 & & & \\
\hline
\multicolumn{7}{c} \hfill MAP$^{+}$  ($w^{-1} \times 0.5 $ (2 yrs),
$\theta_{fwhm}\times 0.75$, $\sigma_{pix} \times 0.94$) \hfill\\
\hline
$w^{-1}/10^{-15}$ & $2.3$ & $2.3$ & $2.3$  &  & & \\
$\ell_s$ & 620 & 460 & 340 & & & \\
\hline
\multicolumn{7}{c} \hfill Planck HFI (first 4 used) \hfill\\
\hline
$\nu_{ch}$(GHz) & 100 & 150 & 220 & 350 & (545) & (857)  \\
$\theta_{fwhm}$ & $14.5^\prime$ & $10^\prime$ & $6.6^\prime$ & $4.1^\prime$ &  ($4.4^\prime$) & ($4.4^\prime$) \\
$\sigma_{pix}/10^{-6}$ & 1.3 & 1.3 & 1.2 & 16 &(77) & (4166)  \\
$w^{-1}/10^{-15}$ & $0.028$ & $0.015$ & $0.005$  & $0.35 $  & & \\
$\ell_s$ & 560 & 800 & 1225 & 1970 & & \\
\hline
\multicolumn{7}{c} \hfill Planck LFI (first 3 used) \hfill\\
\hline
$\nu_{ch}$(GHz) & 100 & 65 & 44 & (30) &  & \\
$\theta_{fwhm}$ & $10^\prime$ & $16^\prime$ & $23^\prime$ &
($34^\prime$) &  & \\
$\sigma_{pix}/10^{-6}$ & 6.2 & 3.7  &2.6 & (1.8) & & \\
$w^{-1}/10^{-15}$ & $0.33$ & $0.29$ & $0.29$  &  & & \\
$\ell_s$ & 810 & 505 & 350 & & & \\
\hline
\end{tabular}
\end{minipage}
\end{table*}

\begin{table*}
\begin{minipage}{120mm}
\caption{Parameter estimation for the 2 models shown. The standard CDM
model has $\Omega_{m}=1$, ${\rm h}=0.5$. The open CDM model has
$\Omega_{m}=0.33$, ${\rm h}=0.6$.  The columns refer to experimental
parameters for a fiducial Long Duration Balloon Experiment (LDB), and
MAP and Planck, with parameters listed in Table 1.  MAP$^{+}$ assumes
a 25\% improvement in beams and 2 years as opposed to 1 year of
observing.  $\Omega_{\Lambda}{\rm h}^2$ is determined with
$\Omega_k{\rm h}^2$ fixed with a prior, and $\Omega_k{\rm h}^2$ is
determined with $\Omega_{\Lambda}{\rm h}^2$ fixed; most other
parameters (except for $\omega_{m}$) are insensitive to fixing either,
or neither.  Values in square brackets indicate the reduced error when
the dominant correlated variable is fixed. Values in circular brackets
indicate what happens when a positivity constraint is imposed. Only
selected cases with significant variations are shown. In the satellite
cases, $\ell_{cut}=2$ was chosen; $\ell_{cut}=12$ was used for the LDB
experiment, with $\ell_{cut}=3$ for DMR which was analyzed with
it. The LDB parameters are based upon observing for ten days with the
TopHat experiment, and assuming 65\% of a $24^\circ$ radius patch will
be usable. Because of the limited sky coverage the LDB
likelihood is combined with the DMR likelihood to constrain low
$\ell$'s. DMR has $\overline{w}^{-1}/10^{-15}=950$. ($\overline{w}$ is
the total weight, eq.(3).) Priors are important for the LDB+DMR
column, make small differences in the MAP column, and essentially none
in the rest. ${\bf P}$ means the prior constraint controls that
parameter's value. Sample values for Helium with no prior are also
shown in curly brackets.}
\label{tab:satparams1} 
\begin{tabular}{|l|l|ll|ll|}
\hline
Param &LDB &  MAP &    MAP$^{+}$ & Planck(LFI) & Planck(HFI) \\
  &  ${\bf P}$ &  no${\bf P}$    &  no${\bf P}$ &no${\bf P}$  & no${\bf P}$ \\
\hline
$\overline{w}^{-1}/10^{-15}$ & 1.5 & 1.5 & .77 & 0.10 & .0033  \\
$f_{sky}$ & .028 & .65 & .65 &  .65 &  .65  \\
\hline
\hline
\multicolumn{6}{|c|}{SCDM MODEL} \\
\hline
\hline
\multicolumn{6}{|c|}{Orthogonal Parameter Combinations within $\eps$} \\
\hline
$\eps <0.01$ & 1/9 &  2/9 &      3/9 &  3/9 & 5/9  \\
$\eps <0.1$ & 4/9 &   6/9 &      6/9 &  6/9 & 7/9   \\
\hline
\multicolumn{6}{|c|}{Single Parameter Errors from Marginalizing Others} \\
\hline
 $\delta \avrg{{\cal C}_\ell}_B^{1/2}/\avrg{{\cal C}_\ell}_B^{1/2}$ & .022 &
.019 (.012) &    .017 &.019 &.015 [.007]\\
 $\delta n_s$  & .13  & .06 (.03)  &    .04 & .01 &.006     \\
 $\delta r_{ts}$  & .89  & .38 (.30)  & .24  & .13  &.09   \\
 $\delta \Omega_b{\rm h}^2/\Omega_b{\rm h_0}^2$& .23 & .09 (.06)  &   .05 &   .016 &.006   \\
 $\delta \Omega_{m}{\rm h}^2/{\rm h_0}^2$& .33 & .18 (.11) & .10  & .04    &  .02  \\
 $\delta \Omega_{\Lambda}{\rm h}^2/{\rm h_0}^2$& .84 & .49 (.35)   & .28  & .14  &  .05  \\
 $\delta \Omega_{hdm}{\rm h}^2/{\rm h_0}^2$ &.25{\bf P}& .07 &    .05&    .04
&     .02  \\
 $\tau_C$  &  .30 &   .22  &   .19 &   .18 & .16  \\
 $\delta Y_{He}/Y_{He}$ & .09{\bf P} & .09{\bf P} $\{1.4\}$  & .09{\bf P} $\{.59\}$ & .08{\bf P} $\{.19\}$   & .07{\bf
P} $\{.10\}$ \\
\hline 
 $\delta \sigma_8/\sigma_8$  &  .28 &  .28 &  .23 &.21 &.18 [.06]\\
 $\delta {\cal P}_\Phi^{1/2}(k_n)/{\cal P}_\Phi^{1/2}(k_n)$ &  .24  &  .24 &  .19 &.17 &.15 [.02] \\
\hline
 $\delta \Omega_{k}{\rm h}^2/{\rm h_0}^2$ & .14 & .07 &        .04  &    .02  & .007  \\
\hline 
 $\delta {\rm h}/{\rm h}$  & .33 &    .19         &    .11  & .06 &.02 \\
\hline
\hline
\multicolumn{6}{|c|}{OPEN CDM  MODEL} \\
\hline
\hline
\multicolumn{6}{|c|}{Orthogonal Parameter Combinations within $\eps$} \\
\hline
$\eps <0.01$ &   2/7 &   2/7  &   2/7 &  3/7 & 5/7    \\
$\eps <0.1$ &   4/7 &   4/7  &   5/7 &  6/7 & 6/7    \\
\hline
\multicolumn{6}{|c|}{Single Parameter Errors from Marginalizing Others} \\
\hline
 $\delta \avrg{{\cal C}_\ell}_B^{1/2}/\avrg{{\cal C}_\ell}_B^{1/2}$ &
.03 &  .02 [.016]  &  .02 &.02 &.016 \\
 $\delta n_s$   &    .10 &    .03  &    .02 & .01 &.003      \\
 $\delta \Omega_b{\rm h}^2/\Omega_b{\rm h_0}^2$ &   .70 &   .17  &   .07 &   .03 &.008  \\
 $\delta \Omega_{m}{\rm h}^2/{\rm h_0}^2$& .41  & .11   & .08  &
.03  &  .006 \\
 $\delta \Omega_{\Lambda}{\rm h}^2/{\rm h_0}^2$ & 1.2  & .31   & .22  & .09  &  .016  \\
 $\tau_C$  &   .24 &   .11 &   .10 &   .07 & .05 \\
\hline
 $\delta \Omega_{k}{\rm h}^2/{\rm h_0}^2$& .17  &        .10 &  .07  &    .03  & .005   \\
\hline
 $\delta {\rm h}/{\rm h}$  &    .76  &    .26  &    .18  & .07 &.013    \\
\hline
\end{tabular}
 \end{minipage}
\end{table*}

\begin{table*}
\begin{minipage}{130mm}
\caption{Parameter estimation for the untilted $\Lambda$CDM
model, with $\Omega_{\Lambda}=0.66$, $\Omega_{k}=0$, ${\rm h}=0.70$,
and the untilted HCDM model, with ${\rm h}=0.50$, 2 species of
neutrinos with the same mass giving $\Omega_{hdm}=0.2$,
$\Omega_{\Lambda}=0$ and $\Omega_{k}=0$. Errors on
$\Omega_{\Lambda}{\rm h}^2$ and $\Omega_k{\rm h}^2$ are independently
determined, as described in the text.  The columns are as in Table 2.}
\label{tab:satparams2} 
\begin{tabular}{{|l|l|ll|ll|}}
\hline
Param &LDB &  MAP &   MAP$^{+}$ & COSA(LFI) & COSA(HFI) \\
  &  ${\bf P}$ &  no${\bf P}$  &   no${\bf P}$ &no${\bf P}$  & no${\bf P}$  \\
\hline
$\overline{w}^{-1}/10^{-15}$ & 1.5 & 1.5 & .77 & 0.10 & .0033  \\
$f_{sky}$ & .028 & .65 & .65 &  .65 &  .65  \\
\hline
\hline
\multicolumn{6}{|c|}{$\Lambda$CDM  MODEL} \\
\hline
\hline
\multicolumn{6}{|c|}{Orthogonal Parameter Combinations within $\eps$} \\
\hline
$\eps <0.01$ & 1/9 &   2/9  &      3/9 &  4/9 & 5/9    \\
$\eps <0.1$ & 4/9 &   6/9 &     6/9 &  7/9 & 7/9   \\
\hline
\multicolumn{6}{|c|}{Single Parameter Errors from Marginalizing Others} \\
\hline
 $\delta \avrg{{\cal C}_\ell}_B^{1/2}/ \avrg{{\cal C}_\ell}_B^{1/2}$ & .021 &  .019 &   .018 &.020 &.017 [.007]\\
 $\delta n_s$   & .20 &    .07  &       .04 & .015 &.010    \\
 $\delta r_{ts}$  &.85 &       .27  (.20) &          .18  & .10  &.08    \\
 $\delta \Omega_b{\rm h}^2/\Omega_b{\rm h_0}^2$ &.36 &   .10  &     .06 &   .02 &.007    \\
 $\delta \Omega_{m}{\rm h}^2/{\rm h_0}^2$&.36 & .12   & .07  &
.03  &  .01 \\
 $\delta \Omega_{\Lambda}{\rm h}^2/{\rm h_0}^2$ &1.0  & .33    & .19  & .09  &  .03    \\
 $\delta \Omega_{hdm}{\rm h}^2/{\rm h_0}^2$ &.16 &    .04&     .03
&     .02  &.006    \\
 $\tau_C$  &.26 &   .18 &     .17 &   .17 & .14  \\
\hline
 $\delta \sigma_8/\sigma_8 $ &  .29 &  .29  &  .24 &.20 &.16 [.08] \\
 $\delta {\cal P}_\Phi^{1/2}(k_n)/{\cal P}_\Phi^{1/2}(k_n)$ &  .28  &  .31   &  .22  &.17 &.15 [.02] \\
\hline
 $\delta \Omega_{k}{\rm h}^2/{\rm h_0}^2$ &.07 & .023 &  .013  &    .006  & .002    \\
\hline
 $\delta {\rm h}/{\rm h}$    &.37 &    .12  &    .07  & .04 &.01  \\
\hline
\hline
\multicolumn{6}{|c|}{HCDM  MODEL} \\
\hline
\hline
\multicolumn{6}{|c|}{Orthogonal Parameter Combinations within $\eps$} \\
\hline
$\eps <0.01$ & 1/9 &  2/9 &      3/9 &  3/9 & 5/9  \\
$\eps <0.1$ & 4/9 &   6/9 &      6/9 &  6/9 & 7/9   \\
\hline
\multicolumn{6}{|c|}{Single Parameter Errors from Marginalizing Others} \\
\hline
 $\delta \avrg{{\cal C}_\ell}_B^{1/2}/\avrg{{\cal C}_\ell}_B^{1/2}$ &
.021 & .017 (.014) &  .016 &.018 &.013 [.008] \\
 $\delta n_s$ &.14   &    .10 (.05)  &    .08 & .04 &.017    \\
 $\delta r_{ts}$  &.73 &       .43 (.28) &         .36  & .20  &.11   \\
 $\delta \Omega_b{\rm h}^2/\Omega_b{\rm h_0}^2$ &.28 &   .12  &   .06 &   .016 &.008    \\
 $\delta \Omega_{m}{\rm h}^2/{\rm h_0}^2$&.32& .14   & .08  & .04  &  .02 \\
 $\delta \Omega_{\Lambda}{\rm h}^2/{\rm h_0}^2$ &.74 & .36 (.23)   & .26  & .15  &  .06   \\
 $\delta \Omega_{hdm}{\rm h}^2/{\rm h_0}^2$ &.43&    .38 (.30)    &    .26
&     .09  &.03   \\
 $\tau_C$  &.33&   .28  &.25 &   .20 & .15  \\
\hline
 $\delta \sigma_8/\sigma_8$   &  .29&  .86 &  .60 &.25 &.14
[.05] \\
 $\delta {\cal P}_\Phi^{1/2}(k_n)/{\cal P}_\Phi^{1/2}(k_n)$  &  .24  &  .27  &  .24  &.19 &.13 [.06] \\
\hline
 $\delta \Omega_{k}{\rm h}^2/{\rm h_0}^2$ &.11 & .05 &  .04  &    .02  & .008    \\
\hline
 $\delta {\rm h}/{\rm h}$ &.27 & .15 &  .12  &    .07  & .02    \\
\hline
\end{tabular}
 \end{minipage}
\end{table*}

\end{document}